\documentclass[reprint,amsmath,amssymb,aps,prl,nofootinbib]{revtex4-2}
\usepackage{amsmath,amsfonts,amstext,amssymb,mathrsfs}
\usepackage{graphicx, xcolor}
\graphicspath{ {Figures/} }
\usepackage[colorlinks=true,linkcolor=navy,urlcolor=navy,
citecolor=navy]{hyperref}
\RequirePackage{color}
\definecolor{navy}{RGB}{0,0,202} 
\usepackage{dcolumn}
\usepackage{bm}
\usepackage{multirow}
\usepackage{comment}
\usepackage{tikz}
\usepackage{mathtools}
\usepackage{float}
\usepackage{slashed}

\usepackage{color}

\usepackage[normalem]{ulem}  

\definecolor{islamicgreen}{rgb}{0.0, 0.56, 0.0}

\begin{document}

\title{Charges, conserved quantities and fluxes in de Sitter spacetime}

\author{Aaron Poole, Kostas Skenderis and Marika Taylor} \affiliation{STAG Research Centre \& Mathematical Sciences, Highfield, University of Southampton, SO17 1BJ Southampton, UK}


\begin{abstract}

We discuss the definition of conserved quantities in asymptotically locally de Sitter spacetimes. 
One may define an analogue of holographic charges at future and past infinity and at other Cauchy surfaces $I_t$ as integrals over the intersection of timelike surfaces $C$ and the Cauchy surface $I_t$. In general, the charges $Q^t$ defined on the Cauchy surface $I_t$ depend on $C$, but if gravitational flux is absent the charges are independent of $C$. On the other hand, if there is a net gravitational flux entering or leaving the spacetime region bounded by $C_1, C_2$ and two Cauchy surfaces then $\Delta Q^t(C_1, C_2)  = Q^t(C_1)-Q^t(C_2)$ changes by the same amount.

\end{abstract}


\maketitle


\paragraph{Introduction} 

Isolated systems with no dynamical gravity have conserved energy and momentum originating from translational invariance. These charges may be obtained from a conserved local energy-momentum tensor via appropriate integrals over a Cauchy surface.  If a system is open there may be flux of energy and momentum through its boundary, 
which is again encoded by the energy-momentum tensor. A prime example of flux is that of electromagnetic radiation originating from a localised source and propagating out to infinity.
When gravity is dynamical  the corresponding symmetries are local and as such the corresponding charges would be zero if there are no boundaries or asymptotic regions.  

\begin{figure}[h] 
\includegraphics[width=0.2\textwidth]{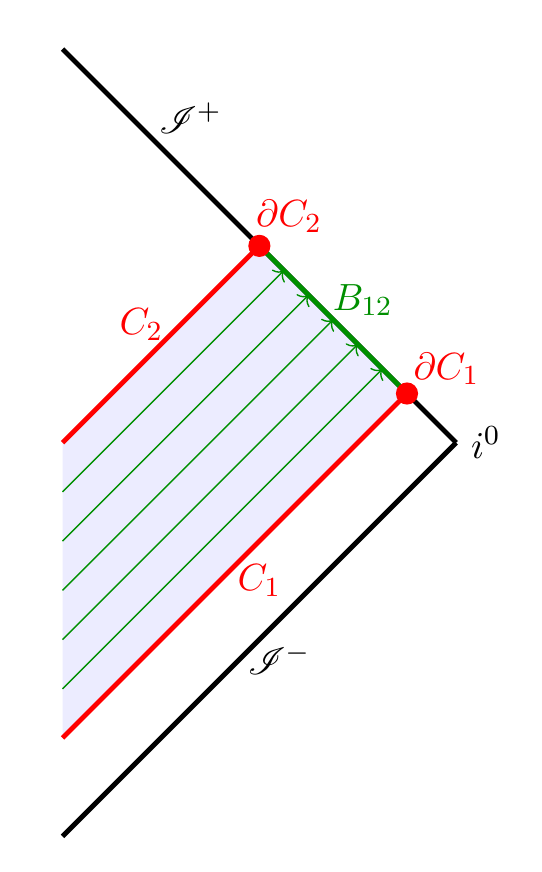}
\caption{Penrose diagram illustrating the flux of gravitational radiation through null infinity, $\mathscr{I}^+$, in an asymptotically flat spacetime. The hypersurfaces $C_1, C_2$ are taken to be null (``instants of retarded time'') although they could also have been chosen to be spacelike. \label{fig:AF}} 
\end{figure}

The study of (conserved) charges in gravitational theories and of gravitational radiation has a long history going back to (at least) \cite{Pirani:1956wr, Trautman:1958zdi, Arnowitt:1959ah, Bondi:1962px, Sachs:1962zza}. Diffeomorphism invariance implies that the charges should be defined by surface integrals at infinity, but  due to the infinite volume of spacetime any such definition requires a subtraction, potentially making the charges ambiguous. Most of the initial work concerned asymptotically flat gravity, and in this context charges were defined relative to flat spacetime. With such asymptotics one may also define outgoing gravitational radiation through null infinity, as illustrated in Fig. \ref{fig:AF}.  In the presence of a cosmological constant the asymptotic structure changes and new issues arise. 

With negative cosmological constant, the anti-de Sitter/Conformal Field Theory (AdS/CFT) correspondence suggests that one should be able to formulate the problem in exactly the same way as when gravity is non-dynamical. Indeed, one can show that such spacetimes are always equipped with a covariantly conserved symmetric tensor which may be used to construct conserved charges. This tensor is identified with the (expectation value of the) energy-momentum tensor of the dual CFT in the AdS/CFT correspondence \cite{deHaro:2000vlm}. Asymptotically locally AdS (AlAdS) spacetimes have a timelike conformal boundary and an associated boundary conformal structure. In the neighborhood of the conformal boundary the metric admits the following Fefferman-Graham form \cite{Fefferman:1985zza},
\begin{equation} \label{eqn:FG1}
ds^2 = \frac{l^2_{AdS}}{\rho^2} \left ( d \rho^2 + g_{ab}(\rho,x) dx^a dx^b \right )
\end{equation}
where $l_{AdS}$ is the curvature radius and in four dimensions:
\begin{equation} \label{eqn:FG2}
g_{ab} = g_{(0) ab} + \rho^2 g_{(2) ab} + \rho^3 g_{(3) ab} + \cdots
\end{equation}
Throughout this Letter for concreteness we specialise to four dimensions. Most of the discussion generalises straightfowardly to other dimensions, modulo issues associated with the holographic conformal anomaly in odd dimensions \cite{Henningson:1998gx}.
Here $g_{(0)}$ is the representative of the boundary conformal structure and the holographic energy momentum tensor is given by \begin{equation} \label{eqn:T}
T_{ab} =\frac{3 l_{AdS}}{16 \pi G} g_{(3)ab}\,,
\end{equation}
where $G$ is Newton's constant.
If the boundary conformal structure admits conformal Killing vectors $\zeta^a$ then the energy-momentum tensor may be used to construct associated conserved charges via a surface integral over a spacelike surface at infinity,
\begin{equation} \label{eqn:Q}
Q_\xi = \int_{\partial C} d s_aT^a_b \zeta^b,
\end{equation} 
where  $\partial C$ is the intersection of the bulk spacelike surface $C$ with the conformal boundary. It was shown in \cite{Papadimitriou:2005ii} using Noether's method and covariant phase space methods that (\ref{eqn:Q}) are the bulk gravitational charges. These charges  are well-defined due to the boundary counterterms 
introduced by holographic renormalisation \cite{Henningson:1998gx, Balasubramanian:1999re, deHaro:2000vlm, Skenderis:2000in}. In fact the counterterms are also required for well-posedness of the variational problem when the boundary conformal structure is fixed \cite{Papadimitriou:2005ii}.

The purpose of this Letter is to address the analogous issues for spacetimes with positive cosmological constant. 
Asymptotically locally de Sitter (AldS) spacetimes share a number of properties with AlAdS spacetimes but there are also important differences. The conformal boundary of AldS spacetimes is spacelike and typically has two causally connected components, one at past infinity and one at future infinity (in contrast the AdS conformal boundary is typically connected\footnote{See \cite{Witten:1999xp} for general results regarding the Euclidean AdS case.}). Near each component the structure of the spacetime is related by analytic continuation to that of a corresponding AlAdS spacetime  \cite{Skenderis:2002wp}, with results analogous to  (\ref{eqn:FG1}), (\ref{eqn:FG2}), (\ref{eqn:T})
(recovering the asymptotic expansion first presented in \cite{Starobinsky:1982mr}).
AldS spacetimes are spatially compact so one cannot define conserved charges as integrals at spatial infinity. Nevertheless, explicit solutions (for example the Schwarzschild dS solution) have integration constants that traditionally are interpreted as the values of conserved charges (like mass and angular momentum). These charges are associated with conformal Killing vectors of the conformal structure at future infinity and may be computed by integrals of the energy momentum tensor at future infinity. However, they are not conserved in the usual sense, {\it i.e.} under time evolution. Instead these charges are conserved under spatial translations if there is no gravitational radiation, as we discuss below.

The discussion of gravitational charges in de Sitter spacetimes goes back to at least  \cite{Abbott:1981ff}, and there has been
renewed interest in this topic since our Universe appears to have positive cosmological constant and gravitational waves have been observed. Recent works include
those of Ashtekar {\it et. al.}  \cite{Ashtekar:2014zfa, Ashtekar:2015lla, Ashtekar:2015lxa, Ashtekar:2015ooa, Ashtekar:2017dlf, Ashtekar:2019khv} (which  builds upon earlier work, \cite{Krtous:2003rw,  Penrose:2011zza}) that discuss charges and radiation for linearised fields on dS, and Chru\'{s}ciel {\it et. al.} \cite{Chrusciel:2016oux, Chrusciel:2020rlz, Chrusciel:2021ttc} where different approaches to the computation of energy and linear fluxes between null cones in dS are considered, see also  \cite{Bishop:2015kay, Date:2015kma, Date:2016uzr, Hoque:2018byx, Kolanowski:2020wfg}. A  number of other approaches to examining radiative bodies in dS can be found in \cite{Szabados:2015wqa,Saw:2017zks, Saw:2017amv,  Szabados:2018erf, Fernandez-Alvarez:2020hsv, Fernandez-Alvarez:2021yog}.
As outlined above, there are many similarities between dS and AdS and works that use the similarity to define charges and fluxes include \cite{Balasubramanian:2001nb, Anninos:2010zf, Anninos:2011jp, Kelly:2012zc, PremaBalakrishnan:2019jvz, Kolanowski:2021hwo,Compere:2019bua, Compere:2020lrt, Fiorucci:2020xto}. 

Our work significantly develops understanding of charges and fluxes in de Sitter. Earlier works were focussed on asymptotically de Sitter spacetimes but our analysis holds for the general class of asymptotically locally de Sitter spacetimes. Restricting to asymptoticially de Sitter does not allow for gravitational waves propagating to the conformal boundary and therefore misses important physical effects. In our work divergences associated with the conformal boundary are treated systematically with holographic renormalisation i.e. covariant boundary counterterms. This approach leads to unambiguous results for changes that are intrinsic to the spacetime rather than defined relative to a reference spacetime and hold for general asymptotically locally de Sitter spacetimes. 

Another key difference with previous works is that we focus on the qualitative new issues that arise in de Sitter, relative to AdS and flat space. One such key difference is the observation that  there are contributions to charges from both endpoints of the timelike surfaces used to define the charges. In previous literature contributions from inner boundaries were often set to be zero by imposing fall off conditions on fields but we do not need to impose any such restrictions. We explain that one needs to consider fluxes through box regions in generic dynamical situations and we give a precise characterisation of this gravitational flux. In particular, we obtain a quantity that is conserved 
under time evolution in the absence of fluxes and equals the flux when gravitational flux is present.

\paragraph{Covariant phase space formalism}

The main tool we use in our analysis is the covariant phase space
\cite{Crnkovic:1986ex, Lee:1990nz, Wald:1993nt, Iyer:1994ys, Iyer:1995kg, Wald:1999wa} (see also the recent works \cite{Harlow:2019yfa, Chandrasekaran:2021vyu} and references therein, and \cite{Barnich:2001jy} for an alternative but equivalent formalism). The key object is 
the $(d-1)$-form {symplectic current} ${\bf{\omega}}$. Starting from a diffeomorphism  covariant {Lagrangian $d$-form} ${\bf{L}}(\psi)$,
the symplectic current is given by
\begin{equation}
{\bf{\omega}} (\psi, \delta_1 \psi, \delta_2 \psi) = \delta_1 \Theta (\psi, \delta_2 \psi) - \delta_2 \Theta (\psi, \delta_1 \psi)\, ,
\end{equation}
where $\Theta$ is the total derivative in the onshell variation of ${\bf{L}}(\psi)$, $\delta {\bf {L}} = d {\Theta}$
($\psi$ denotes the metric and other fields). A crucial property of ${\bf {\omega}}$ is that it is closed onshell,
\begin{equation}
d \omega=0\, .
\end{equation} 

Consider now onshell configurations $\psi$ whose variations $\delta \psi$ are also onshell. 
For a diffeomorphism generated by $\xi$ the corresponding Hamiltonian $H_{\xi}$ is given by
\begin{equation} \label{eqn:H}
\delta H_{\xi} 
= \int_C {\bf \omega} (\psi, \delta \psi, {\cal L}_{\xi} \psi)
\end{equation}
with $C$ a $(d-1)$-dimensional slice, which for many applications is a Cauchy surface,
and ${\cal L}_{\xi}$ is the Lie derivative. 
Given such a diffeomorphism  we can associate a $(d-1)$-form Noether current, 
${\bf J} (\xi) = \Theta (\psi, {\cal L}_{\xi} \psi) - i_{\xi} {\bf L}$,
where $i_{\xi}$ is the interior product. 
The current
${\bf J}$ is closed onshell and is thus locally exact: ${\bf J} (\xi) = \text{d} {\bf Q} (\xi)$
with $(d-2)$-form Noether charge ${\bf Q}$, the variation of $H_{\xi}$ may be written as
\begin{equation}
\delta H_{\xi} = \int_{\partial C} \left ( \delta {\bf Q }(\xi) - i_{\xi} \Theta(\psi, \delta \psi ) \right )
\end{equation}
where $\partial C $ is the boundary of $C$. 

One can integrate this equation to obtain $H_{\xi}$ iff the 
right hand side is a total variation, requiring
\begin{equation}
\int_{\partial C} i_{\xi} \Theta (\psi, \delta \psi) = \delta \int_{\partial C} i_{\xi} {\bf B} (\psi )
\end{equation}
Note that one can also express the condition for existence of $H_{\xi}$ in terms
of $[ \delta_1, \delta _2 ] H_{\xi} = 0$:
\begin{equation} \label{eq: integrability} 
\int_{\partial C} i_{\xi} \omega (\psi, \delta_1 \psi, \delta_2 \psi)  = 0. 
\end{equation}
The difference between Hamiltonians at two different $C$s is given by 
\begin{equation} \label{eq: Hamiltonian_difference}
\left. \delta H_{\xi} \right\vert_{\partial C_2} - \left. \delta H_{\xi} \right\vert_{\partial C_1} = - \int_{B_{12}} {\bf \omega} (\psi, \delta \psi, {\cal L}_{\xi} \psi)
\end{equation}
where $B_{12}$ is a hypersurface with boundary $\partial C_1 \sqcup \partial C_2$. If (\ref{eq: integrability}) is satisfied by virtue of $\omega|_{B_{12}} = 0$, then equation (\ref{eq: Hamiltonian_difference}) tells us that $H_{\xi}$ is independent of the slice $C$.
We now review the application of this formalism to AlAdS spacetimes, before discussing AldS spacetimes.

\paragraph{Asymptotically locally anti-de Sitter:}

This case was fully analysed in \cite{Papadimitriou:2005ii}. The slice $C$ in (\ref{eqn:H}) is a bulk spacelike surface that ends on the boundary of AlAdS. Considering Dirichlet conditions\footnote{Other boundary conditions may be obtained by adding finite boundary terms that implement this change of boundary condition and tracking their contributions, see also \cite{Compere:2020lrt,Fiorucci:2020xto}. \label{ft:BC}}, where the conformal class $[g_{(0)}]$ is kept fixed, and a $[g_{(0)}]$ that admits conformal Killing vectors, the integrability criterion (\ref{eq: integrability}) is automatically satisfied by virtue of $\left. \omega(g_{\mu \nu}, \delta g_{\mu \nu}, \mathcal{L}_{\xi} g_{\mu \nu}) \right\vert_{B_{12}}=0$, where here $\xi$ is an asymptotic conformal Killing vector (ACKV), {\it i.e.} it approaches the conformal Killing vectors $\zeta$ at the boundary, see appendix B of \cite{Papadimitriou:2005ii} for the specific fall-off conditions)
Then the Wald Hamiltonian integrates to (\ref{eqn:Q}).

\begin{figure}[h]
\includegraphics[width=0.2\textwidth]{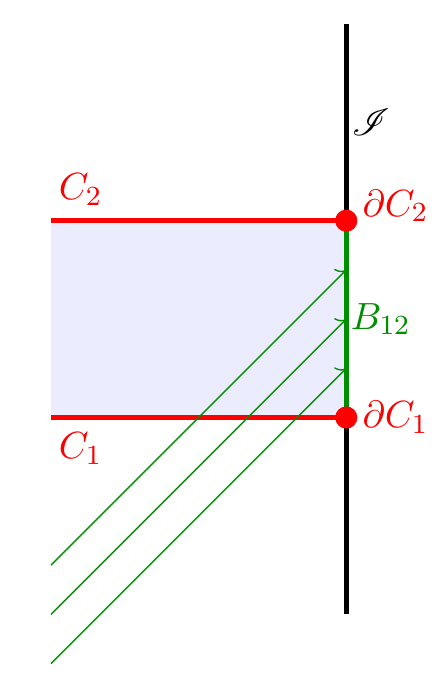}
\caption{Illustrating flux in an AlAdS spacetime. If flux arrives in $B_{12}$, the gravitational charge at $C_2$ will differ from that in $C_1$. The conformal boundary $\mathscr{I}$  is a timelike hypersurface and the $C_1, C_2$ are spacelike.  } \label{fig:AdS}
\end{figure}

When there is gravitational radiation from the interior, it  will reach the conformal boundary at finite proper time. As a result, radiative AdS spacetimes 
have a time-dependent boundary conformal structure, as was confirmed recently in \cite{Poole:2018koa}. In such cases there are no conserved charges because the boundary conformal structure does not admit conformal Killing vectors. However, if the gravitational flux arrives in finite time intervals, so that the boundary geometry outside these intervals admits conformal 
Killing vectors, there will be piece-wise constant gravitational charges, and the difference in their values between two instances of time will account for the gravitational flux arriving at the conformal boundary, see  Fig. \ref{fig:AdS}. 
More generally, if the spacetime admits a limit 
where conserved quantities exist, then one can discuss the flux of such quantities away from that limit. An example is the Robinson-Trautmann solution which we will discuss below in the context of $\Lambda >0$.

\paragraph{Asymptotically locally de Sitter:} 

\begin{figure}[h]
\includegraphics[width=0.25\textwidth]{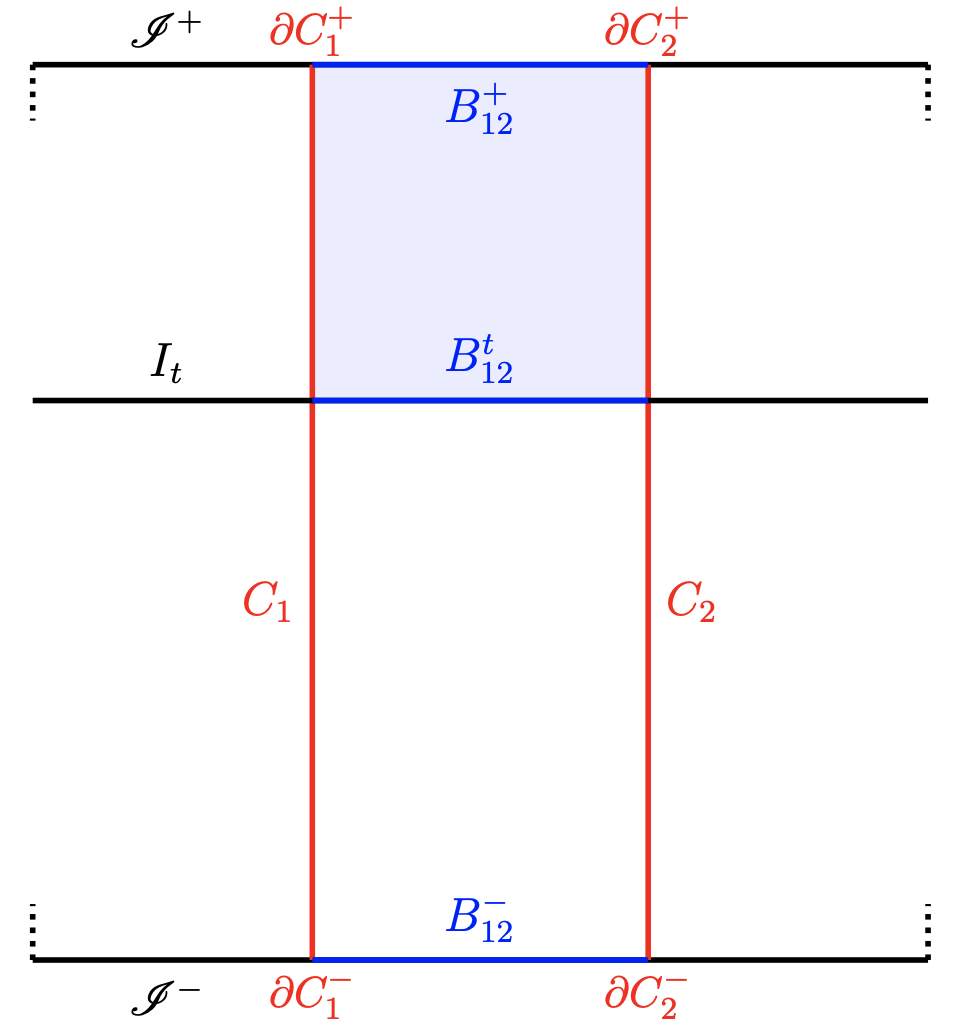}
\caption{Timelike $C_1,C_2$ intersecting spacelike conformal boundaries $\mathscr{I}^{\pm}$  in AldS spacetime.}
 \label{fig: dS} 
\end{figure}

We now consider AldS spacetimes with Dirichlet boundary conditions, {\it i.e.} we keep fixed a conformal class at each end. Different boundary conditions may be implemented as discussed in footnote \ref{ft:BC}.
One may adapt straightforwardly all steps of $\Lambda<0$ case to $\Lambda >0$; in most formulae this amounts to differences in signs. However, there are important conceptual differences. The analogue of the surface $C$ in (\ref{eqn:H}) we used in the AlAdS case is now timelike, and because the conformal boundary has two components, one at future infinity and another at past infinity, such a $C$ will have two ends.
Let us consider the case where both past and future infinity admit conformal Killing vectors and
a hypersurface  $C$ that extends from future infinity to past infinity. Let $\xi$ be a bulk vector that is an ACKV near both past and future conformal infinity, approaching a boundary conformal Killing vector  $ \zeta_{\pm}^b$ at $\mathscr{I}^\pm$.Then
\begin{equation} \label{eqn: ds_charges1}
H_{\xi}[C] =Q_\xi^+ - Q^-_\xi
\end{equation}
where $Q^\pm_\xi[C]=\int_{\partial C^{(\pm)}} ds_a T_{(\pm) b}^a  \zeta_{\pm}^b$; $T_{(\pm) b}^a$ is the holographic energy momentum tensor, and  $\partial C^{\pm}$  are the two ends of $C$. Now consider two such surfaces $C_1$ and $C_2$, see Fig. \ref{fig: dS}. Then 
(\ref{eq: Hamiltonian_difference}) implies
\begin{equation} \label{eq: Charge_difference}
\Delta Q_\xi^+(C_1,C_2) = \Delta Q_\xi^-(C_1,C_2) =0
\end{equation}
where $\Delta Q_\xi^\pm(C_1,C_2) = Q^\pm_\xi[C_1] - Q^\pm_\xi[C_2]$. The first equality in (\ref{eq: Charge_difference}) follows from  $\left. \omega(g_{\mu \nu}, \delta g_{\mu \nu}, \mathcal{L}_{\xi} g_{\mu \nu}) \right\vert_{B^\pm_{12}}=0$ and the second from applying Stokes' theorem to $\Delta Q_{\xi}^{\pm}$ independently at $B_{12}^{\pm}$, using the explicit formulae for $Q^{\pm}_{\xi}$. It follows that while the values of the charges may change from 
$\mathscr{I}^-$ to $\mathscr{I}^+$, their difference is zero. Recall that $\mathscr{I}^\pm$ are Cauchy surfaces and consider any other Cauchy surface $I_t$. Then as we now argue, $\Delta Q_\xi^t = \Delta Q_\xi^-$, where $\Delta Q_\xi^t(C_1, C_2) = Q^t_\xi[C_1] - Q^t_\xi[C_2]$, and
$Q^t_\xi[C]= \int_{C \cap I_t} ds_a 2 \pi_{b}^a  \xi^b$, with $\pi_{ab}$ the conjugate momentum of the induced metric at $I_t$ (here, for simplicity, we assume  that we deal with pure gravity).
In other words, $ \Delta Q_\xi^t$ is conserved under time evolution, {\it i.e.} independent of the Cauchy surface $I_t$. 
{To see this, note that since $\omega$ is closed, its integral over the boundary of the region enclosed by $C_1, C_2, I_t, \mathscr{I}^-$ is zero, so $\Delta Q_\xi^-$ must be equal to $\Delta Q^t$ plus any flux passing through $B_{12}^t$}. However, there cannot be such flux, since any flux would eventually reach future infinity and as a result the conformal class at $\mathscr{I}^+$ would not admit conformal Killing vectors. Since we assume that future infinity admits conformal Killing vectors, there cannot be gravitational flux in the interior, and we find $\Delta Q_\xi^- = \Delta Q_\xi^t$. {So while $Q_\xi^t$ may depend in $t$, $\Delta Q_\xi^t=0$ for all $t$}. One may {also} check that known solutions such as Kerr-de Sitter have $\Delta  Q^+_\xi=0$.

In our discussion we consider timelike slices $C$ that extend from the future infinity to past infinity. One may also consider hypersurfaces that start and end at future infinity. Such $C$ may be obtained by using the portion of the hypersurface $C_1$  from future infinity till $I_t$, then join $B_{12}^t$ and return to future infinity using (the time-reverse of) $C_2$. 

Let us now consider the effects of radiation, \textit{i.e.} we consider the case when $\mathscr{I}^+$ admits no conformal Killing vectors. In this case the same argument implies that $\Delta Q_\xi^+$ will differ from $\Delta Q_\xi^t$ by the amount of net radiation in the region bounded by $\mathscr{I}^+, C_1, C_2$ and $I_t$. In many previous discussions the contribution from $I_t$ is either ignored or argued to be zero by postulating suitable decay rates in the deep interior.  Assuming $\Delta Q_\xi^t=0$, any non-zero $\Delta Q_\xi^+$ will be due to gravitational radiation at $\mathscr{I}^+$, captured explicitly via the \textit{future flux formula} \cite{Anninos:2010zf, Compere:2020lrt, Kolanowski:2021hwo}
\begin{equation} \label{eq: future_flux_gen}
\Delta Q_{\xi}^+(C_1, C_2) =  - \int_{B_{12}^{+}} \bm{F}_{\xi_{(0)}} 
\end{equation}
where 
\begin{equation} \label{eq: future_flux_exp}
 \bm{F}_{\xi_{(0)}} =  \left(-\frac{1}{2} \sqrt{g_{(0)}} T^{ab} \mathcal{L}_{\xi_{(0)}} g^{(0)}_{ab} \right) \bm{\epsilon}_3.
\end{equation}

Using our formalism, we can also consider the case of $\Delta Q_\xi^t \neq 0$, \textit{i.e.} account for radiation passing through the interior hypersurface $I_t$. The \textit{past flux formula} is 
\begin{equation} \label{eq: past_flux_gen}
 \Delta Q_{\xi}^t(C_1, C_2) = - \int_{B_{12}^{t}} \bm{G}_{\xi}  
\end{equation}
where 
\begin{equation}
 \bm{G}_{\xi} = \left( - \sqrt{\gamma}  \pi^{ab} L_{\xi} \gamma_{ab} \right) \bm{\epsilon}_3.
\end{equation}
In summary, the effect of gravitational radiation is to spoil the conservation of spacelike separated charges. Combining equations (\ref{eq: future_flux_gen}) and (\ref{eq: past_flux_gen}), we see that the explicit difference  is given by 
\begin{equation}
\Delta Q^+ -  \Delta Q^t = - \int_{B_{12}^{+}} \bm{F}_{\xi_{(0)}} + \int_{B_{12}^{t}} \bm{G}_{\xi}.
\end{equation}

\paragraph{Robinson-Trautman}

To conclude we discuss an example of a class of radiative dS solutions: the Robinson-Trautman (RT) family. RT \cite{Robinson:1960zzb} describes an isolated system which relaxes to equilibrium by radiating excess energy via gravitational waves, becoming Schwarzschild dS. The solution exists for any value of the cosmological constant, and here we discuss the $\Lambda>0$ case. 
The metric  is 
\begin{equation} \label{eq: RT_metric}
ds^2 = - \Phi_{\Lambda} du^2 -2 du dr +2 r^2 P^{-2} d\zeta d\bar{\zeta}\,,
\end{equation}
where $P = P(u, \zeta, \bar{\zeta})$ and
\begin{equation}
\Phi_{\Lambda} = \Delta \ln P - 2r (\ln P)_{,u}-\frac{2m}{r} - \frac{\Lambda}{3}r^2\,.
\end{equation}
$\Delta = 2 P^2 \partial^2/\partial \zeta \partial \bar{\zeta}$ and $P$ satisfies the \textit{Robinson-Trautman} equation:
\begin{equation} \label{eq: RT_equation}
(\ln P)_{,u} + \frac{1}{12M} \Delta \Delta( \ln P ) = 0.
\end{equation}


Using results from \cite{Bakas:2014kfa}, a representative of the conformal class at $\mathscr{I}^+$ is
\begin{equation}
ds_{(0)}^2=g_{(0) ab} dx^a dx^b = dy^2 + \frac{6}{\Lambda} \hat{P}^{-2} d\zeta d\bar{\zeta}
\end{equation}
where $\hat{P}$ denotes the boundary value of $P$. This metric generically possesses no conformal Killing vectors and thus we cannot define charges. However, as $y \to \infty$ the solution approaches Schwarzschild dS which has an isometry related to mass, and therefore we may discuss its flux.
Let us consider a vector field which in this limit is associated with 
\textit{Bondi ``time" translation} along the boundary (note that it is spacelike). The vector generating these translations is
\begin{equation}
\xi_{(0)}^a \partial_a = \frac{\hat{P}_0}{\hat{P}} \partial_y 
\end{equation}
where $\hat{P}_0=1+ \zeta \bar{\zeta} /2$ is the limit of $\hat{P}$ as $y \to \infty$.
The normalisation of this vector relates to the RTdS metric (\ref{eq: RT_metric}) needing to undergo a coordinate transformation to be expressible in Bondi gauge \cite{Aranha:2013rj}. Exploiting results from \cite{Bakas:2014kfa}, 
one can show that 
\begin{equation} \label{eq: Bondi_mass_RT} 
Q^+_{\xi} = \frac{2m}{\kappa^2} \int_{\partial C^{+}} \left( \frac{P_0}{P} \right)^3  \, d\mu_0 
\end{equation}
where $d\mu_{0}$ is the area element on the unit $S^2$. This quantity is analogous to the \textit{Bondi mass}, $\mathcal{M}_{B}$, of RT in the asymptotically flat setting \cite{Bondi:1962px}. It has been previously observed \cite{Chrusciel:1992cj, Chrusciel:1992rv, Bakas:2014kfa} that the quantity defined in equation (\ref{eq: Bondi_mass_RT}) is monotonically decreasing regardless of the sign of $\Lambda$. Our construction thus illustrates explicitly how the Bondi mass of the asymptotically flat RT solution arises in the current context. 


Since we have not used an asymptotic conformal Killing vector to construct $\mathcal{M}_{B}$, we should expect to discover additional flux through $B_{12}^{+} \subset \mathscr{I}^+$ when comparing the difference in Bondi mass between the hypersurfaces $C_1$ and $C_2$. Applying equations  (\ref{eq: future_flux_gen}) and (\ref{eq: future_flux_exp}) to the RT solution, we find the following equation for the flux through $B_{12}^{+}$:
\begin{equation}
\Delta Q_\xi^+ =\left. \mathcal{M}_B \right\vert_{\partial C^{+}_1} - \left. \mathcal{M}_B \right\vert_{\partial C^{+}_2}  = - \int_{B_{12}^{+}} \bm{F}_{\xi_{(0)}}
\end{equation}
where
\begin{equation}
\bm{F}_{\xi_{(0)}} = -\frac{2m}{\kappa^2}  \partial_y \left(\frac{P_0}{P} \right)^3 \sqrt{\mathring{\sigma}} \bm{\epsilon}_3
\end{equation}
 and $\mathring{\sigma}_{AB} dx^A dx^B = 2P_0^{-2} d\zeta d\bar{\zeta}$ is the metric on the unit round $S^2$. This flux formula gives 
\begin{equation}
\left. \mathcal{M}_B \right\vert_{\partial C^{+}_1} - \left. \mathcal{M}_B \right\vert_{\partial C^{+}_2} \leq 0 \,. \label{ineq}
\end{equation} 
by monotonicity of the expression defined in equation (\ref{eq: Bondi_mass_RT}) \cite{Chrusciel:1992cj, Chrusciel:1992rv}. We note that the existence of this monotonically decreasing charge is unlikely to be a feature of all AldS spacetimes, where charges may obey flux-balance laws similar to those proposed in \cite{Compere:2019bua}. However, this analysis proves that all metrics in the RT(A)dS family possess such a charge.

It would be interesting to compute $\Delta Q_\xi^t$ and confirm monotonicity in $u$. For this we would need to know 
the form of Bondi time translation vector field in the deep interior of the RT spacetime. 
We leave this computation to future work.

\paragraph{Conclusions} We have discussed the definition of charges in asymptotically locally de Sitter spacetimes. We show that the difference of charges $\Delta Q^t_\xi(C_1,C_2)$ defined using timelike surfaces $C_1$ and $C_2$  at the Cauchy surface $I_t$ is zero if gravitational flux is absent.  If there is gravitational flux the corresponding quantities differ by the net amount of flux. We illustrated the latter with the example of Robinson-Trautmann dS and derived a bound on the flux \eqref{ineq}.
 It would be interesting to exemplify our discussion with further explicit examples, investigate possible connections with work on gravitational wave memory \cite{Chu:2015yua, Bieri:2015jwa, Kehagias:2016zry, Chu:2016qxp, Tolish:2016ggo, Chu:2016ngc, Hamada:2017gdg, Bieri:2017vni}, and analyse possible relevance to present and future gravitational wave observations. Covariant phase space has been instrumental in formulating in generality and obtaining a deep understanding of the first law of black hole thermodynamics  \cite{Wald:1993nt, Iyer:1994ys}. Thermodynamics of black holes with de Sitter asymptotics  is still poorly understood and we believe the framework we develop here and the issues we uncover are important in this context as well.

\vspace{3mm}

\begin{acknowledgments}
{\em Acknowledgments.} 
 KS and MT acknowledge support from the Science and Technology Facilities Council (Consolidated Grant “Exploring the Limits of the Standard Model and Beyond”). AP is supported by an EPSRC fellowship.
 
\end{acknowledgments} 
 
\bibliography{Letter}

\end{document}